\documentclass{PoS}
\usepackage{amsmath}

\renewcommand{\thefigure}{\arabic{figure}}

\newcommand\as{\alpha_{\mathrm{S}}}

\newcommand\f[2]{\frac{#1}{#2}}
\def\ep{\epsilon}

\def\wp{\widetilde P}

\def\beq{\begin{equation}}
\def\eeq{\end{equation}}
\def\beeq{\begin{eqnarray}}
\def\eeeq{\end{eqnarray}}
\def\cm{{\cal M}}
\def\bom#1{{\mbox{\boldmath $#1$}}}
\def\bom#1{{\mbox{$\mathbf{#1}$}}}
\def\to{\rightarrow}

\newcommand{\la}{\langle}
\newcommand{\ra}{\rangle}

\def\nn{\nonumber}

\def\ID{1 \kern -.45 em 1}

\def\sp{{\bom {Sp}}}
\def\ket#1{|{#1}\ra}

\def\cmbar{{\overline {\cal M}}}
\def\Mbar{{\overline M}}
\def\imc{{\bom I}^{(1)}_{m\,C}}

\def\dmc{{\Delta}^{(1)}_{m\,C}}
\def\imctwo{{\bom I}^{(2)}_{m\,C}}

\def\cbet0{b_0}

\title{Factorization violation in the multiparton collinear limit}

\ShortTitle{Factorization violation in the multiparton collinear limit}

\author{Stefano Catani \\
INFN, Sezione di Firenze and Dipartimento di Fisica e Astronomia,
Universit\`a di Firenze, I-50019 Sesto Fiorentino, 
Florence, Italy \\
E-mail: \email{stefano.catani@fi.infn.it}}

\author{Daniel de Florian \\
Departamento de F\'\i sica and IFIBA, FCEYN, Universidad de Buenos Aires, \\
(1428) Pabell\'on 1 Ciudad Universitaria, Capital Federal, Argentina \\
E-mail: \email{deflo@df.uba.ar}}

\author{\speaker{Germ\'an Rodrigo} \\
Instituto de F\'{\i}sica Corpuscular, UVEG - Consejo Superior de 
Investigaciones Cient\'{\i}ficas, \\ 
Parc Cient\'{\i}fic, E-46980 Paterna (Valencia), Spain \\
E-mail: \email{german.rodrigo@csic.es}}

\abstract{We present an all-order generalized factorization formula for QCD scattering 
amplitudes in kinematical configurations where two or more momenta of the external partons 
become collinear. The singular behaviour of the scattering amplitudes in the collinear limit 
is encoded by collinear splitting matrices.
In the space-like region and beyond the tree level, 
the collinear splitting matrices
depend also on the momenta and quantum numbers 
of the non-collinear partons, thus breaking strict collinear factorization. 
Although the factorization breaking contribution partly cancels for squared amplitudes, 
due to its one-loop absorptive origin, remaining effects at high perturbative orders 
have implications on the non-abelian structure of logarithmically-enhanced terms in perturbative 
calculations 
and on various factorization
issues of mass singularities. 
}

\FullConference{Loops and Legs in Quantum Field Theory - 11th DESY Workshop on Elementary Particle Physics,\\
		April 15-20, 2012\\
		Wernigerode, Germany}

\begin{document}

\section{Introduction}

A central topic in QCD and, more generally, in gauge field theories
is the structure of infrared (virtual, soft and 
collinear) singularities of the perturbative scattering amplitudes. 
The divergent or singular behaviour of scattering amplitudes 
is described by corresponding {\em factorization formulae} 
and it is captured by factors that have a high
degree of universality or, equivalently, a minimal process dependence
(i.e. a minimal dependence on the specific scattering amplitude).

In the following, we consider the factorization formulae describing 
the singular behaviour of QCD amplitudes in kinematical configurations 
where two or more external-parton momenta become collinear.
In the case of two collinear partons at the tree level,
the collinear-factorization formula for QCD {\em squared amplitudes}
was first derived in Ref.~\cite{Altarelli:1977zs}.
The corresponding factorization for QCD {\em amplitudes} was introduced in 
Refs.~\cite{Berends:1987me, Mangano:1990by}.
At the tree level, the multiple collinear limit of three, four or more partons 
has been studied 
\cite{Campbell:1997hg, Catani:1998nv, DelDuca:1999ha, Birthwright:2005ak,Catani:2012ri}
for both amplitudes and squared amplitudes.
In the case of {\em one-loop} QCD amplitudes, collinear factorization was
introduced in Refs.~\cite{Bern:1993qk, Bern:1995ix, Bern:1998sc, 
Kosower:1999rx},
by explicitly treating the collinear limit of two partons.
Explicit, though partial, results for the triple collinear limit of one-loop
amplitudes were presented in Ref.~\cite{Catani:2003vu}.
The two-parton collinear limit of {\em two-loop} amplitudes
was explicitly computed in Refs.~\cite{Bern:2004cz, Badger:2004uk}.
The structure of collinear factorization of higher-loop amplitudes 
is discussed in Refs.~\cite{Kosower:1999xi, Catani:2011st, Forshaw:2012bi}.

The singular collinear factors are customarily expected to depend {\em only} on the
momenta and quantum numbers (flavour, colour, spin) of the collinear partons, 
with no dependence on the external {\em non-collinear} partons. 
This feature of collinear factorization is denoted as 
strict collinear factorization, and it is generally assumed to be valid in the 
calculation of cross-sections in hadron--hadron collisions from 
the convolution of universal (i.e., process independent) parton 
distribution functions with the hard-scattering cross-section. 
Strict collinear factorization, however, is violated 
beyond the tree-level for amplitudes 
in {\em space-like} collinear configurations~\cite{Catani:2011st}.
The violation is originated \cite{Catani:2011st, Forshaw:2012bi}
by long-wavelength absorptive contributions
(such as those produced by 
Coulomb--Glauber gluons~\cite{Ralston:1982pa, Catani:1985xt,Bonciani:2003nt,
Forshaw:2006fk, Aybat:2008ct, Bauer:2010cc,DelDuca:2011ae})
that causally disconnect initial-state and final-state interactions,
thus limiting the factorization features due to colour coherence.
Owing to the absorptive (`imaginary') origin of the 
violation of strict factorization, the effect is partly canceled 
at the level of {\em squared amplitudes}.
Indeed, such a cancellation is complete up to the next-to-leading
order (NLO). Nonetheless, strict factorization is violated at higher orders. 
This challenges the validity of the factorization theorem of mass 
(collinear) singularities~\cite{Collins:1989gx, Mitov:2012gt} and related issues in the 
context of the factorization of transverse-momentum dependent distributions
\cite{Bomhof:2004aw, Bacchetta:2005rm, Rogers:2010dm}, and it can produce
logarithmically-enhanced radiative 
corrections~\cite{Forshaw:2006fk,Catani:2011st}
to hard-scattering processes in hadron--hadron collisions.

\section{Generalized collinear factorization at all orders}

A set $\{ p_1, \dots, p_m \}$ of $m$ ($m \geq 2$) parton momenta 
approaches the multiparton collinear limit when they become parallel. 
In this limit all the parton subenergies
\beq
s_{i \ell}=(p_i+p_\ell)^2~, 
\quad \quad {\rm with} \;\;\;\;\; i,\ell \in C= \{\,1,\dots,m \,\} \;\;, 
\label{subenergies}
\eeq
are of the {\em same} order and 
vanish {\em simultaneously}~\cite{Campbell:1997hg,Catani:1998nv}.  
The collinear direction is defined through the light-like vector 
\beq
\label{ptilm}
{\wp}^\mu = 
p_{1,m}^\mu - \frac{p_{1,m}^2 \; n^\mu}{2 \, n \cdot p_{1,m}}~,
\eeq
where $n^\mu$ is an auxiliary light-like vector ($n^2=0$), 
which parametrizes how the collinear limit is approached,
and $p_{1,m}=p_1 + \dots + p_m$.  
The longitudinal-momentum fractions $z_i$ are 
\beq
\label{zim}
z_i = \frac{n \cdot p_i}{n \cdot {\wp}} =
\frac{n \cdot p_i}{n \cdot p_{1,m}} \;\;, \quad \;\;\; 
i \in C \;\;,
\eeq
and they fulfill the constraint $\sum_{i=1}^m z_i =1$, 
with $p_i^\mu \to z_i \, {\wp}^\mu (i \in C)$ in the collinear limit.
According to our notation, $p_i^\mu$ is the {\em outgoing} momentum in the
scattering amplitude, and the time component (`energy') $p_i^0$ is positive
(negative) for a final-state (initial-state) parton. 
In the time-like (TL) collinear region all the parton subenergies in 
Eq.~(\ref{subenergies}) are positive and $1> z_i > 0$; in all the other
kinematical configurations, 
we are dealing with the space-like (SL) collinear region.
Therefore, in the TL case, {\em all} the collinear partons 
are either final-state partons or initial-state partons.
In the SL case, at least one collinear parton is in the initial state 
and, necessarily, one or more partons are in the final state.
The SL collinear limit is typically encountered by considering initial-state
radiation in hadron collision processes.

%%%%%%%%%%%%%%%%%%%%%%%%%%%%%%%%%%%%%%%%%%%%%%%%%%%%%%%%%%%%%%%%%%%%%%%%%%%%%%%%

In the multiparton collinear limit the matrix element $\cm$ 
of the scattering process with $n$ external partons ($n>m$)
fulfills the {\em all-order} generalized factorization 
formula~\cite{Catani:2011st}
\beq
\label{factallL}
\ket{\cm} \;
\simeq \;\sp(p_1,\dots,p_m;{\widetilde P};p_{m+1},\dots,p_n) 
\;\;\ket{\cmbar} \;\;,
\eeq
where $\sp$ is the all-order splitting matrix, which captures 
the dominant singular behaviour in the multiparton collinear limit, 
%$(1/\sqrt{s})^{m-1}$ mod $(\ln^k s)$,
and $\cmbar = \cm(\wp,p_{m+1},\dots,p_n)$ is the reduced matrix element, 
which is obtained from the original matrix element $\cm$
by replacing the $m$ collinear partons $A_1,\dots,A_m$ 
with a single parton $A$, which carries the momentum
$\wp$. The matrix element and the splitting matrix 
satisfy the perturbative (loop) expansion:
\beeq
\label{loopexmren}
\cm &=& 
\cm^{(0)} + \cm^{(1)} + \cm^{(2)}  + \dots~, \\
\label{loopexspren}
\sp &=& 
\sp^{(0)} + \sp^{(1)} + \sp^{(2)}  + \dots~,  
\eeeq
where the superscripts $(k)$ ($k=0,1,2,\dots$) refer to the
order (number of loops) of the perturbative expansion.
%Since renormalization {\em commutes} with the collinear limit,
%we implicitly assume renormalized quantities 
%in Eqs.~(\ref{loopexmren}), (\ref{loopexspren}) 
%and in the following. 
The splitting matrix is expected to be 
universal and process independent (strictly factorized), 
namely, it should depend on the momenta and quantum 
numbers (flavour, colour, spin) of the external collinear partons only.
Nonetheless, according to  
Eq.~(\ref{factallL}), the splitting matrix can also acquire a dependence 
on the {\em non-collinear} partons. 
Strict collinear factorization holds at the tree-level (in both the TL and
SL regions):
\beeq
\label{loopexsp}
\sp^{(0)}(p_1,\dots,p_m;{\widetilde P};p_{m+1},\dots,p_n) = 
\sp^{(0)}(p_1,\dots,p_m;{\widetilde P})~,
\quad ({\rm TL \,\; and \,\; SL \,\; coll. \,\; lim.})~, 
\eeeq
and it also holds (because of colour coherence) 
in the TL 
collinear region to all orders:
\beq
\label{colsptlallL}
\sp(p_1,\dots,p_m;{\widetilde P};p_{m+1},\dots,p_n) =
\sp(p_1,\dots,p_m;{\widetilde P})~, 
\quad ({\rm TL \,\; coll. \,\; lim.})~.
\eeq
%Because of colour coherence,  
%however, strict collinear factorization holds in the TL 
%collinear region to all orders 
%\beq
%\label{colsptlallL}
%\sp(p_1,\dots,p_m;{\widetilde P};p_{m+1},\dots,p_n) =
%\sp(p_1,\dots,p_m;{\widetilde P})~, 
%\quad ({\rm TL \,\; coll. \,\; lim.})~,
%\eeq
%and at the tree-level in the SL region (through Eq.~(\ref{colsptlallL})
%also in the TL)
%\beeq
%\label{loopexsp}
%\sp(p_1,\dots,p_m;{\widetilde P};p_{m+1},\dots,p_n) = 
%\sp^{(0)}(p_1,\dots,p_m;{\widetilde P}) + \dots~,
%\quad ({\rm TL \,\; and \,\; SL \,\; coll. \,\; lim.})~. 
%\\eeeq
In the SL region, strict collinear factorization is 
violated~\cite{Catani:2011st} at one-loop and 
%at each 
higher-loop orders. 

We illustrate the violation of strict collinear factorization by mainly
considering the infrared (IR) divergent part of the splitting matrix.
The IR structure of 
$\sp$
%the splitting matrix 
is not independent \cite{Catani:2003vu, Bern:2004cz, Becher:2009qa, Dixon:2009ur,
Ahrens:2012qz}
of the IR structure of the
%corresponding multiloop 
QCD amplitude $\cm$.
%~\cite{Catani:1998bh,csdip,Aybat:2006wq}.
Using dimensional regularization in $d=4-2\ep$ space-time dimensions,
%To {\em all-orders}, 
the all-order matrix element $\cm$ 
%(and the reduced matrix element $\overline \cm$) 
fulfills the 
IR recursion 
relation~\cite{Catani:1998bh, Sterman:2002qn, Aybat:2006wq, Dixon:2008gr,
Becher:2009cu, Gardi:2009qi}
\beq
\label{mallir}
\ket{\cm } = {\bom I}_M(\ep) \;
\ket{\cm } +  \;\ket{\cm^{\,{\rm fin.}}} 
\;\;,
\eeq
where the operator 
%${\bom I}_M(\ep)$ 
${\bom I}_M(\ep)= {\bom I}_M^{(1)}(\ep) + {\bom I}_M^{(2)}(\ep) + \dots$
(with perturbative 
coefficients ${\bom I}_M^{(k)}(\ep)$)
is IR divergent, 
%with perturbative 
%coefficients ${\bom I}_M^{(k)}(\ep)$ $(k=1, 2, \ldots)$, 
while the matrix element term 
%$\cm^{\,{\rm fin.}}$ 
$\cm^{\,{\rm fin.}} = \cm^{(0)} + \cm^{(1)\,{\rm fin.}} + \cm^{(2)\,{\rm fin.}}
+ \dots$
is IR finite and 
its first contribution in the perturbative expansion is the 
complete tree-level matrix element $\cm^{(0)}$ in Eq.~(\ref{loopexmren}).
%($\cm^{\,{\rm fin.}} = \cm^{(0)} + \cm^{(1)\,{\rm fin.}} + \cm^{(2)\,{\rm fin.}}$).
An expression analogous to Eq.~(\ref{mallir}) holds for $\overline \cm$ 
and the corresponding IR operator ${\bom I}_{\overline{M}}(\ep)$.
The {\em all-order} splitting matrix also fulfills a recursion 
relation~\cite{Catani:2011st}:
\beq
\label{spfindef}
\sp = {\bf V}(\ep) \;\,\sp^{\,{\rm fin.}} 
\;\,{\bf {\overline V}}^{\,-1}(\ep)
= \left[ \,1 -  {\bf {\overline V}}(\ep) \;{\bf V}^{-1}(\ep)
\,\right] \,\sp + {\bf {\overline V}}(\ep) \;\,\sp^{\,{\rm fin.}} 
\;\,{\bf {\overline V}}^{\,-1}(\ep)~,
\eeq
where the first term in the perturbative expansion of
the 
%all-order 
IR finite splitting matrix $\sp^{\,{\rm fin.}}$ is 
the tree-level splitting matrix $\sp^{(0)}$ ($\sp^{\,{\rm fin.}} = \sp^{(0)} 
+ \sp^{(1)\,{\rm fin.}} + \sp^{(2)\,{\rm fin.}} + \ldots$), and 
\beq
\label{vm1}
{\bf V}^{-1}(\ep) = 1 - {\bom I}(\ep)~,
\qquad {\bf {\overline V}}^{-1}(\ep) = 1 - {\bom {\overline I}}(\ep)~,
\eeq
where ${\bom I}$ and ${\bom {\overline I}}$ are obtained 
from the collinear limit of the 
%corresponding 
IR operators 
${\bom I}_M$  and ${\bom I}_{\Mbar}$, respectively. 
Up to two loops, the expansion of the coefficient of the first term in 
the right-hand side of Eq.~(\ref{spfindef}) reads 
$1 -  {\bf {\overline V}}(\ep) \;{\bf V}^{-1}(\ep) = \imc(\ep) +
\imctwo(\ep) + {\cal O}(\as^3)$, where
\beeq
\label{i1mc}
\imc(\ep) &=& {\bom I}^{(1)}(\ep) - {\bom {\overline I}}^{(1)}(\ep)~, \\
\label{i2mc}
\imctwo(\ep) &=& {\bom I}^{(2)}(\ep) - {\bom {\overline I}}^{(2)}(\ep)
+ {\bom {\overline I}}^{(1)}(\ep) 
\left( \,{\bom I}^{(1)}(\ep) - {\bom {\overline I}}^{(1)}(\ep) \,
\right) \;\;.
\eeeq

\begin{figure}
\begin{center}
\includegraphics[width=5cm]{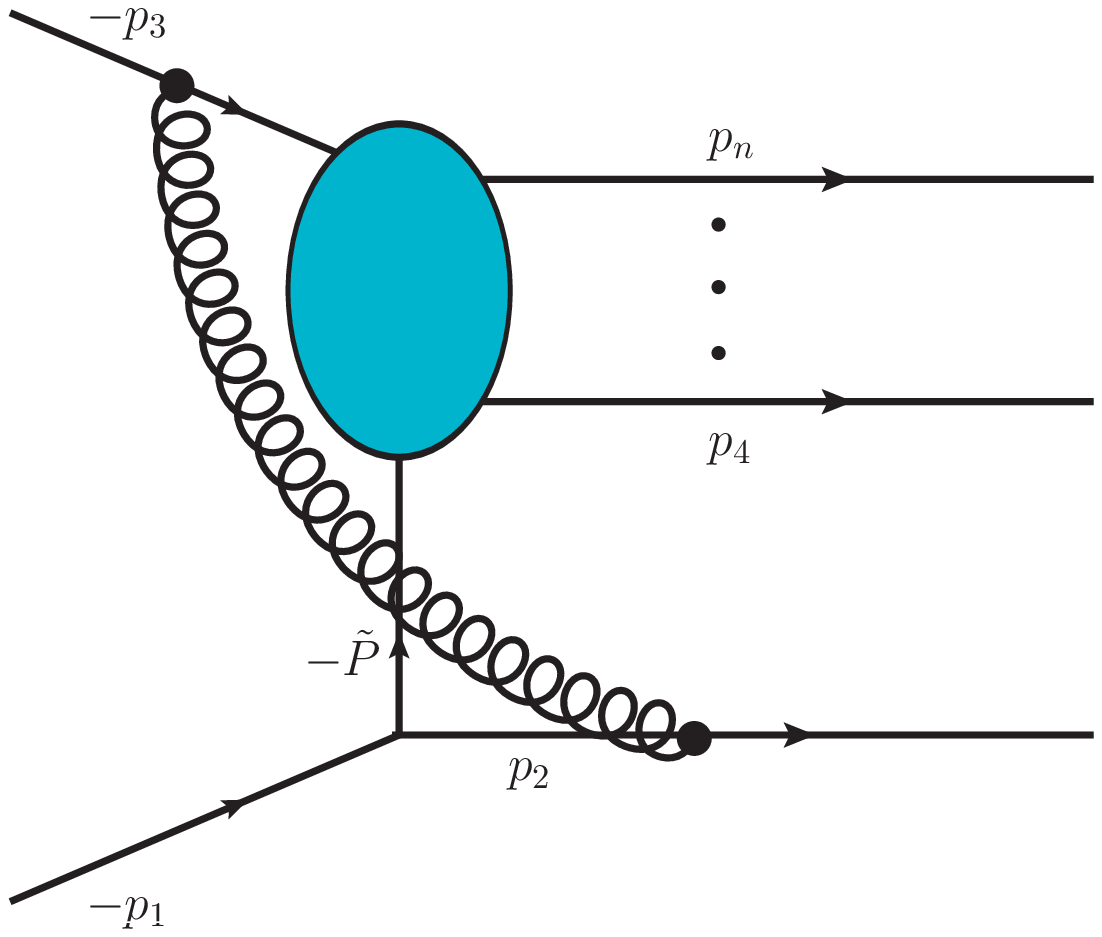} \qquad
\includegraphics[width=5cm]{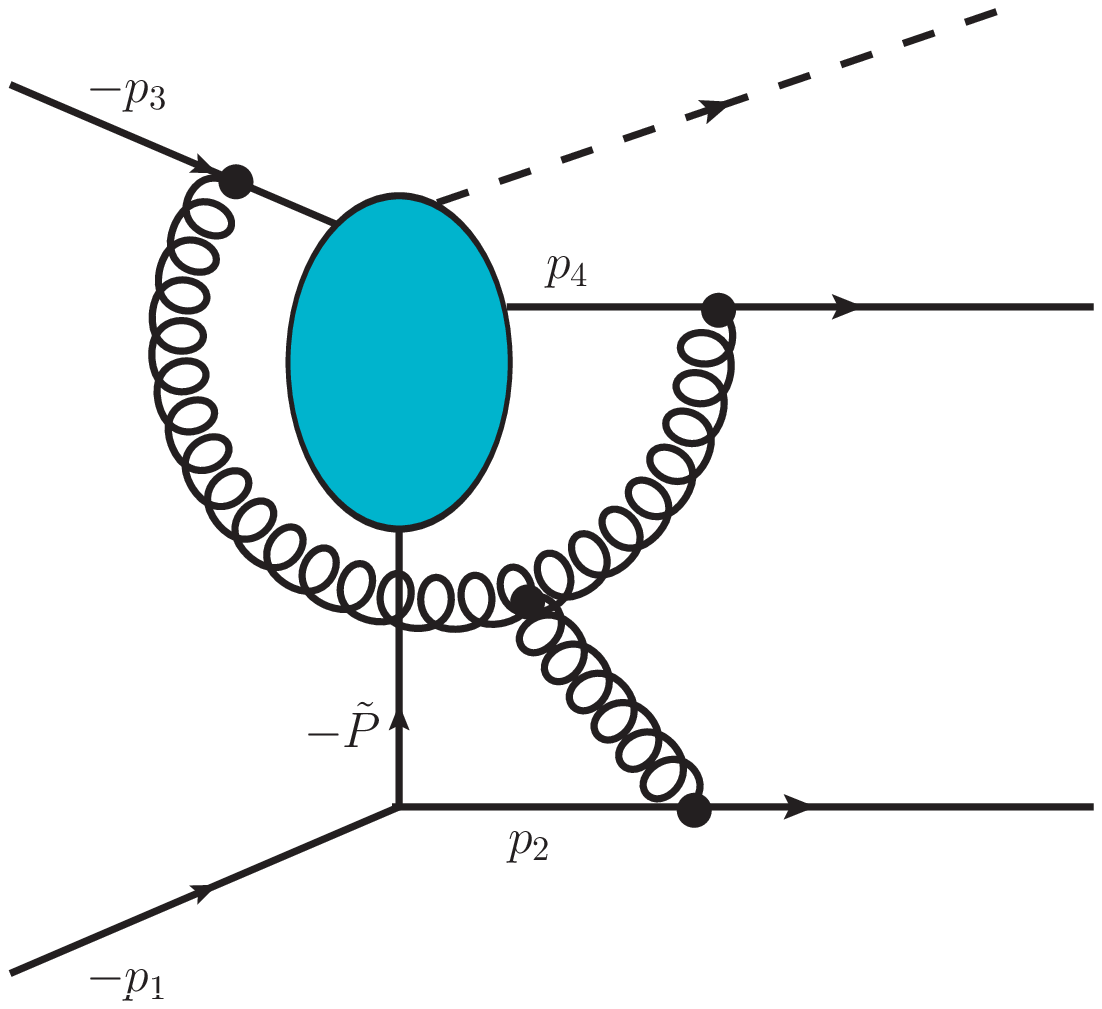}
\caption{Two-parton factorization breaking correlations at one-loop 
in hadron--hadron collisions (left).
Three-parton factorization breaking correlations at two-loops, $n \ge 4$ 
QCD partons (right).}
\label{colourcorrelations}
\end{center}
\end{figure}

The IR operators in Eqs.~(\ref{i1mc}) and (\ref{i2mc}) have been 
calculated 
%in the TL and SL regions
in Ref.~\cite{Catani:2011st} starting from the 
%well 
known 
IR structure of scattering amplitudes  
to two-loop order~\cite{Catani:1998bh, Aybat:2006wq}. 
In the SL collinear region,
the operator $\imc(\ep)$, which describes the IR divergent part 
of the one-loop splitting matrix $\sp^{(1)}$,
contains factorization breaking contributions that are proportional 
to the {\it anti-Hermitian} operator
\beq
\label{del1}
\dmc(\ep) = \f{\as(\mu^2)}{2\pi} \,\;\f{i \pi}{\ep} \;
\sum_{\substack{i \,\in C \\ j \,\in NC}} 
{\bom T}_i \cdot {\bom T}_j \;\Theta(-z_i) \;{\rm sign}(s_{ij}) \;\;,
\eeq
where $NC=\{m+1, \ldots, n\}$ denotes the set of non-collinear partons, and
${\bom T}_k$ is the colour-charge matrix of the $k$-th parton
(we are using the general colour space notation of Ref.~\cite{csdip}). 
The operator in Eq.~(\ref{del1}) embodies colour correlations 
between collinear and non-collinear partons that are produced by 
the non-Abelian Coulomb phase, and thus they violate strict collinear 
factorization. 
In the two-parton 
collinear case,
these colour correlations 
are illustrated in Fig.~\ref{colourcorrelations}~(left).
Note that 
the two-parton one-loop SL splitting matrix is known 
to all orders in $\epsilon$~\cite{Catani:2011st} and, therefore, the result
of Ref.~\cite{Catani:2011st} is 
not limited to the treatment of Coulomb-Glauber gluon effects 
to leading IR 
%(or logarithmic) 
accuracy. For three or more 
collinear partons, the IR finite part of $\sp^{(1)}$ is unknown 
(the  
explicit calculation for three collinear partons is in progress), 
but it also contains factorization breaking contributions 
similar to those in Eq.~(\ref{del1}).
%Through colour conservation, Eq.~(\ref{del1})
%can equivalently be written in terms of colour correlations
%involving the initial state partons $1$ and $m+1$
%and the parent parton, namely a colour factor ${\bom T}_{m+1}\cdot 
%({\bom T}_1 - {\bom T_{\widetilde P}})$~\cite{Forshaw:2012bi}.

At two loops, new factorization breaking terms appear 
in the SL region through the operator~\cite{Catani:2011st} 
\beeq
\label{d22mc}
{\Delta}_{m \,C}^{(2;\,2)}(\ep) &=& \left(\f{\as(\mu^2)}{2\pi}\right)^2  
\left( \,- \, \f{1}{2 \,\ep^2} \right) \;\pi \; f_{abc} \,
\sum_{i \in C} \;\sum_{\substack{j, k \,\in NC \\ j \,\neq\, k}} 
\,T_i^a \,T_j^b \, T_k^c \;\Theta(-z_i) \;
{\rm sign}(s_{ij}) \;\Theta(-s_{jk}) \nn \\
&\times& \ln\left(- \,\f{ s_{j\wp} \; s_{k\wp}}{s_{jk} \,\mu^2} -i0 \right)~,
\eeeq
which contributes (to $\imctwo$ and) to the IR divergent part of $\sp^{(2)}$.
The operator ${\Delta}_{m \,C}^{(2;\,2)}$
includes both {\it Hermitian} and {\it anti-Hermitian} 
contributions, and it embodies three-parton correlations involving 
one collinear and two non-collinear partons 
(see Fig.~\ref{colourcorrelations}~(right)). The Hermitian 
part also depends on the size of the non-collinear momenta. The 
anti-Hermitian part, which depends only on the sign of the partons subenergies, 
can be rewritten~\cite{Catani:2011st} 
in terms of correlations between two collinear and one non-collinear 
partons.
%The two-parton collinear configuration ($m=2$) 
%requires to consider scattering amplitudes with $n\ge 4$ QCD partons
%(if $n=4$ there has to be additional colourless external legs, as shown 
%in Fig.~\ref{colourcorrelations} (right), otherwise the corresponding reduced 
%matrix element $\overline \cm$ vanishes and Eq.~(\ref{d22mc}) would 
%not lead to any contribution, this is true also for $n-m=2$).

The breaking of strict collinear factorization in the SL region 
beyond the tree-level is due to absorptive contributions originated 
from the fact that colour coherence is limited by 
%the causality 
%structure of long-range gauge interactions distinguishing 
%initial-state from final-state interactions.
causality, which distinguishes initial-state from final-state interactions.
Lepton-hadron DIS is, however, a special case since all the non-collinear
coloured partons are produced in the final state, and thus there
are no initial-state interactions between collinear and non-collinear
partons. The one-loop and two-loop SL multiparton splitting matrices 
explicitely calculated in Ref.~\cite{Catani:2011st} effectively 
take a strictly-factorized form in DIS. 

\section{Squared amplitudes and cross-sections}

The all-order singular behaviour of the squared matrix element $|\cm|^2$ 
(summed over the colours and spins of the external partons)
in a generic kinematical configuration of $m$ collinear partons 
is obtained by squaring the generalized factorization formula in 
Eq.~(\ref{factallL}):
\beq
\label{factallm2}
|\cm|^2 \;
\simeq \la \, \cmbar \,
 | \, \; {\bf P}(p_1,\dots,p_m;{\widetilde P};p_{m+1},\dots,p_n) 
\;\;\ket{\cmbar} \;\;,
\eeq
where the matrix ${\bf P}$ is the square of the all-order splitting 
matrix $\sp\,$,
\beq
\label{mpdef}
{\bf P} \equiv
\left[\, \sp \,\right]^\dagger \,\sp~,
\eeq
and it satisfies a loop expansion similar to Eq.~(\ref{loopexspren}), 
with perturbative coefficients ${\bf P}^{(k)}$ $(k=0,1,2, \ldots)$. 

The splitting matrix in hadron--hadron collision processes 
breaks strict collinear factorization beyond the tree-level. However, 
owing to their absorptive origin, the factorization breaking effects 
partly cancel for {\it squared amplitudes}, and thus a violation of 
collinear factorization in physical observables is not directly implied. 
Factorization breaking contributions 
can cancel either in the squared splitting matrix ${\bf P}$,  
after taking the expectation value with the reduced matrix 
element in Eq.~(\ref{factallm2}), 
or even among different partonic processes with different 
number of external partons contributing to the same physical 
cross-section at a given order.

The tree-level collinear matrix ${\bf P}^{(0)}$ is obviously strictly 
factorized also in the SL region. 
The divergent part of ${\bf P}^{(1)}$ is strictly factorized 
because the factorization breaking operator
in Eq.~(\ref{del1}) is anti-Hermitian. In the two-parton 
collinear limit also the complete ${\bf P}^{(1)}$, including 
the IR finite part, is strictly factorized~\cite{Catani:2011st}, 
in spite of the fact that $\sp^{(1)}$ violates collinear factorization.
Similarly, the three-parton factorization breaking correlations 
appearing in the two-parton splitting matrix ${\bf \sp}^{(2)}$ 
(Fig.~\ref{colourcorrelations}~(right)) survive in ${\bf P}^{(2)}$ 
but only for processes with $n\ge 5$ QCD partons~\cite{Catani:2011st}. 

The expectation value of the two-loop operator 
${\widetilde {\Delta}}_{P}^{(2)}(\ep) = 
{\widetilde {\Delta}}_{C}^{(2)}(\ep) + {\rm h.c.}$,
which gives the (IR dominant) factorization
breaking contribution to the squared splitting matrix 
${\bf P}^{(2)}$ for the two-parton collinear limit~\footnote{The operator 
${\widetilde {\Delta}}_{C}^{(2)}(\ep)$, 
which is computed in Ref.~\cite{Catani:2011st}, is analogous to 
the multiparton collinear operator ${\Delta}_{m \,C}^{(2;\,2)}(\ep)$ 
but it includes also the subdominant 
$1/\ep$ poles and some finite terms to all orders in $\ep$, which are known 
in the two-parton collinear configuration.},
onto the reduced matrix element $\cmbar$ is
\beeq
\label{vev}
&&  \!\!\!\!\!\! \!\!\!\!\!\!\!\!\! \la \, \cmbar \, |
\left( \sp^{(0)} \,\right)^\dagger
\;{\widetilde {\Delta}}_{P}^{(2)}(\ep) \;\sp^{(0)} \;
\, \ket{\cmbar} =
 \la \, \cmbar^{(0)} \,
 |  
\left( \sp^{(0)} \,\right)^\dagger 
\;{\widetilde {\Delta}}_{P}^{(2)}(\ep) \;\sp^{(0)} 
\;\ket{\cmbar^{(0)}} 
\nn \\
&&\;\; +\; \left[ \la \, \cmbar^{(1)} \,
 | \, 
\left( \sp^{(0)} \,\right)^\dagger 
\;{\widetilde {\Delta}}_{P}^{(2)}(\ep) \;\sp^{(0)} 
\;\ket{\cmbar^{(0)}} + {\rm c.c.} \; \right] + {\rm higher \;orders}
\;\;.
\eeeq
Although ${\widetilde {\Delta}}_{P}^{(2)}$ is not vanishing,
its lowest-order expectation value 
(i.e., the first term on the right-hand side of Eq.~(\ref{vev})) 
vanishes in {\em pure} QCD~\cite{Forshaw:2012bi} (i.e., if the lowest-order
reduced matrix element $\cmbar^{(0)}$ is produced by tree-level QCD
interactions). Note however that this contribution would 
be non-vanishing, for instance, for tree-level 
quark--quark scattering produced by electroweak interactions (with 
CP-violating electroweak couplings and/or finite width of the $Z$ 
and $W^{\pm}$ bosons), or if the tree-level QCD scattering is 
supplemented with one-loop (pure) QED radiative corrections.

Therefore, as a matter of principle, it remains true that the operator  
${\widetilde {\Delta}}_{P}^{(2)}$ explicitly uncovers {\em two-loop}
QCD effects that lead to violation of strict collinear factorization at the
squared amplitude level. Moreover, the second term in the 
right-hand side of Eq.~(\ref{vev}) is not vanishing and contributes
to the SL collinear limit of {\it three-loop} QCD squared amplitudes, 
together with the factorization breaking effect produced by ${\bf P}^{(3)}$ and
highlighted independently in Refs.~\cite{Catani:2011st, Forshaw:2012bi}.
In the case of scattering amplitudes with $n=5$ QCD partons, the colour
correlation structure of these two factorization 
breaking contributions at three loops is analogous to the commutator structures
that were found in the N$^4$LO computation of super-leading logarithms
in `gaps--between--jets' cross sections~\cite{Forshaw:2006fk}.

\section{Summary}

Collinear factorization has been generalized to all orders for kinematical configurations
where two or more external partons become collinear, and explicit results on one-loop and 
two-loop (and three-loop) amplitudes for both the two-parton and multiparton collinear limits
have been presented \cite{Catani:2011st}. In the space-like region strict factorization is 
%broken 
violated
beyond the tree-level for scattering amplitudes. For squared amplitudes the 
strict-factorization breaking terms partly cancel, but the remaining effects still 
lead to logarithmically-enhanced contributions at high perturbative orders and challenge the 
validity of factorization theorems of mass (collinear) singularities.

{\it Acknowlegdements:} This work is supported by REA Grant Agreement PITN-GA-2010-264564 (LHCPhenoNet), 
by UBACYT, CONICET, ANPCyT, INFN, INFN-MICINN agreement AIC-D-2011-0715, 
MICINN (FPA2007-60323, FPA2011-23778 and CSD2007-00042 CPAN) and 
GV (PROMETEO/2008/069).

\end{document}